\pdfoutput=1

\documentclass[ amsmath, amssymb, aps, prd, twocolumn, lengthcheck, sort&compress, showpacs ]{revtex4-1}

\usepackage{amsmath}
\usepackage{amssymb}
\usepackage{graphicx}
\usepackage{dsfont}
\usepackage{dcolumn}
\usepackage{units}
\usepackage{bm}
\usepackage{wasysym}

\usepackage[usenames]{color}
\usepackage[colorlinks=true,linkcolor=blue,citecolor=blue,urlcolor=blue]{hyperref}

      \newcommand{\conjg}[1]{\ensuremath{\hspace{1pt}\overline{\hspace{-1pt}#1\hspace{-1pt}}}\hspace{1pt}}

      \newcommand{\be}{\begin{equation}}
      \newcommand{\ee}{\end{equation}}

      \def\Slash#1{\setbox0=\hbox{$#1$} % set a box for #1
      \dimen0=\wd0 % and get its size
      \setbox1=\hbox{/} \dimen1=\wd1 % get size of /
      \ifdim\dimen0>\dimen1 % #1 is bigger
      \rlap{\hbox to \dimen0{\hfil/\hfil}} % so center / in box
      #1 % and print #1
      \else % / is bigger
      \rlap{\hbox to \dimen1{\hfil$#1$\hfil}} % so center #1
      / % and print /
      \fi}

      \def\longlongrightarrow{
      \relbar\joinrel\relbar\joinrel\relbar\joinrel\relbar\joinrel\rightarrow}
      
      \def\longlonglonglongrightarrow{
      \relbar\joinrel\relbar\joinrel\relbar\joinrel\relbar\joinrel\relbar\joinrel\relbar\joinrel\relbar\joinrel\relbar\joinrel\relbar\joinrel\rightarrow}

\begin{document}

\title{Unified description of hadron-photon and hadron-meson scattering \\ in the Dyson-Schwinger approach}

         \author{Gernot~Eichmann}
%         \affiliation{Institut f\"{u}r Theoretische Physik I, Justus-Liebig-Universit\"at Giessen, D-35392 Giessen, Germany  }
         \email{gernot.eichmann@theo.physik.uni-giessen.de}

         \author{Christian~S.~Fischer}
         \affiliation{Institut f\"{u}r Theoretische Physik, Justus-Liebig-Universit\"at Giessen, D-35392 Giessen, Germany}

         \date{\today}

         \begin{abstract}
         We derive the expression for the nonperturbative coupling of a hadron to two external currents
         from its underlying structure in QCD. Microscopically, the action of each current is resolved
         to a coupling with dressed quarks. The Lorentz structure of the currents is arbitrary and thereby
         allows to describe the hadron's interaction with photons as well as mesons in the same framework.
         We analyze the ingredients of the resulting four-point functions and explore their potential
         to describe a variety of processes relevant in experiments. Possible applications include
         Compton scattering, the study of two-photon effects in hadron form factors,
         pion photo- and electroproduction on a nucleon, nucleon-pion or pion-pion scattering.

         \end{abstract}


         \maketitle

%%%%%%%%%%%%%%%%%%%%%%%%%%%%%%%%%%%%%%%%%%%%%%%%%%%%%%%%%%%%%%%%%%%%%%%%%%%%%%%%%%%%%%%%%%%%%%%%%%%%%%%%%%%%%%%%%%%%%%%%%%%%%%%%%%%%%%%%%%%%%%%%%%%%%%%%%%%%%%%%%%%%%%%%%%%%%%%
%%%%%%%%%%%%%%%%%%%%%%%%%%%%%%%%%%%%%%%%%%%%%%%%%%%%%%%%%%%%%%%%%%%%%%%%%%%%%%%%%%%%%%%%%%%%%%%%%%%%%%%%%%%%%%%%%%%%%%%%%%%%%%%%%%%%%%%%%%%%%%%%%%%%%%%%%%%%%%%%%%%%%%%%%%%%%%%

        \section{Introduction}

            Pion- and photon-induced reactions with the nucleon have been the main source
            of experimental information on the structure and properties of light hadrons.
            In addition to the data collected in $N\pi$ scattering, pion photo- and electroproduction experiments
            have provided new insight in the electromagnetic properties of nucleon resonances such as the $\Delta(1232)$
            or $N(1440)$~\cite{Pascalutsa:2006up,Drechsel:2007sq,Klempt:2009pi,Tiator:2011pw,Aznauryan:2011qj}.
            Moreover, virtual Compton scattering in the low-energy region determines the nucleon's generalized polarizibilities
            and, at higher energies, allows to resolve its spatial and spin structure~\cite{Drechsel:2002ar,Belitsky:2001ns,Burkardt:2008jw},
            thereby providing a connection to the underlying substructure in Quantum Chromodynamics (QCD).
            The Compton scattering amplitude also determines the magnitude of two-photon effects
            in the extraction of nucleon form factors~\cite{Arrington:2011dn}.

            Depending on the studied momentum range, different theoretical approaches are employed to describe these reactions.
            In deeply virtual Compton scattering (DVCS), the factorization property of QCD allows to disentangle perturbative and nonperturbative
            contributions and model the soft parts of the process by generalized parton distributions~\cite{Goeke:2001tz,Diehl:2003ny,Belitsky:2005qn}.
            Pion-nucleon and pion photoproduction on the other hand are analyzed by unitarized chiral perturbation theory~\cite{Bernard:2007zu},
            phenomenological partial-wave analyses~\cite{Anisovich:2005tf,Arndt:2006bf,Drechsel:2007if}
            or dynamical coupled-channel models~\cite{Shklyar:2004ba,Surya:1995ur,Matsuyama:2006rp,Chen:2007cy,Huang:2011as}.
            Here, the dominant $s-$ and $t-$channel poles are usually implemented via potentials
            and the scattering matrices are generated by iterating Bethe-Salpeter equations at a purely hadronic level.
            Features such as unitarity and analyticity of the $S-$matrix,
            electromagnetic gauge invariance and crossing symmetry are desired but not always realized to full extent, see Ref.~\cite{Huang:2011as} for a discussion of this issue.
            Nevertheless, these approaches provide important theoretical input to the experimental data analyses
            and are necessary to reconstruct baryon form factors from the experimental cross sections.

\begin{figure*}[t]
                    \begin{center}

                    \includegraphics[scale=0.32]{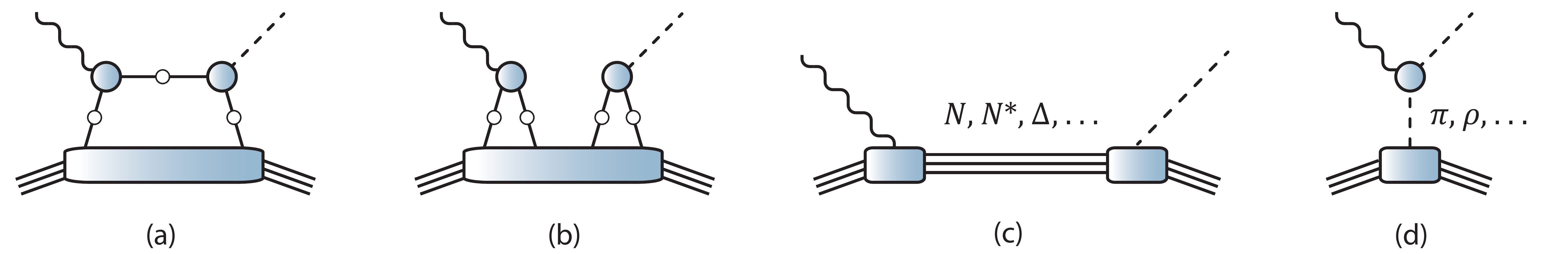}
                    \caption{(Color online) Various contributions to the four-point function,
                             here in the case of pion photoproduction on the nucleon, that must be recovered
                             in a microscopic description: handbag (a) and cat's-ears diagrams (b);
                             $s-$channel and, via crossing, $u-$channel nucleon resonances (c);
                             $t-$channel meson exchange (d).   } \label{fig:preliminary}

                    \end{center}
        \end{figure*}

            A desirable alternative would be a nonperturbative calculation of such scattering amplitudes within QCD itself,
            i.e., by resolving the underlying dynamics in terms of quarks and gluons.
            Typical diagrams that a microscopic description should recover in various kinematical ranges are shown in Fig.~\ref{fig:preliminary}
            for the case of pion photo- and electroproduction. Microscopically, the pion and the photon couple to dressed quarks.
            This leads to handbag diagrams (Fig.~\ref{fig:preliminary}a) which, in the analogous case of virtual Compton scattering,
            become dominant at large photon virtualities, as well as so-called cat's-ears diagrams (Fig.~\ref{fig:preliminary}b)
            that comprise two active quarks. However, the relevant contributions in the low-energy regions are others,
            namely processes that involve $s-$channel nucleon resonances and $t-$channel meson exchange
            (Fig.~\ref{fig:preliminary}c and \ref{fig:preliminary}d). A comprehensive description should resolve those diagrams in
            terms of their quark and gluon substructure as well. The purpose of this paper is to derive such expressions in a systematic way.

            We work in the framework of Dyson-Schwinger equations (DSEs) of QCD~\cite{Alkofer:2000wg,Fischer:2006ub,Roberts:2007jh}.
            They interrelate QCD's Green functions and thereby provide access to nonperturbative
            phenomena such as dynamical chiral symmetry breaking and confinement.
            In combination with covariant bound-state equations, i.e., Bethe-Salpeter equations (BSEs) and Faddeev equations,
            they present a comprehensive framework for studying hadron properties from QCD.
            The approach allows to probe the substructure of hadrons at all momentum scales
            and all quark masses. It has been applied to compute a variety
            of hadron observables such as meson and baryon masses, wave functions,
            leptonic decay constants, elastic and transition form factors;
            see~\cite{Maris:2003vk,Roberts:2007jh,Eichmann:2009zx,Blank:2011qk,Chang:2011vu} and references therein.

            Building upon that, a systematic and nonperturbative procedure
            to construct hadron matrix elements from the underlying Green functions
            is the 'gauging--of--equations' method~\cite{Haberzettl:1997jg,Kvinikhidze:1998xn,Kvinikhidze:1999xp}.
            The basic idea is to couple an external current such as a photon
            to all internal building blocks to arrive at a quark-level description
            of hadronic matrix elements, for example, electromagnetic form factors.
            If all ingredients satisfy appropriate Ward-Takahashi identities (WTIs),
            electromagnetic gauge invariance is satisfied by construction.
            The method has been used for computing nucleon form factors both in the
            quark-diquark model~\cite{Oettel:1999gc} and from the covariant three-body equation~\cite{Eichmann:2011vu}.
            The scope of the approach is however more general: the same formalism can be exploited
            to derive generalized parton distributions from their quark-gluon substructure~\cite{Kvinikhidze:2004dy}, and
            it has been recently applied to obtain the scattering amplitudes
            for $N\pi$ scattering and pion electroproduction at a hadronic level, i.e.,
            from chiral effective field theories~\cite{Blankleider:2010mg,Haberzettl:2011zr}.

            In the following we will use the gauging procedure to resolve the coupling
            of a hadron to \emph{two} external currents from its substructure in QCD.
            The hadronic scattering matrices obtained in that way are constructed from
            QCD's Green functions and hadron wave functions which, in turn,
            must be determined beforehand from their DSEs and bound-state equations.
            As a consequence, all hadron resonances shown in Fig.~\ref{fig:preliminary}
            can be expressed through diagrams that involve only the degrees of freedom in the QCD Lagrangian.
            Since the method does not discriminate between the scattering on mesons or baryons,
            and the involved currents can have any desired Lorentz structure and thus describe photons as well as mesons,
            one arrives at a unified theoretical framework for a variety of reactions including
            nucleon Compton scattering ($N\gamma^\star  \rightarrow N\gamma^\star $), pion-nucleon scattering ($N\pi \rightarrow N\pi $),
            pion photo- and electroproduction ($ N\gamma^\star \rightarrow  N\pi$), or $\pi\pi$ scattering.
            We will show in Section~\ref{sec:mesons-in-rl} that a rainbow-ladder truncation,
            i.e., a gluon-exchange interaction between quark and antiquark,
            recovers the diagrams that were employed in Refs.~\cite{Bicudo:2001jq,Cotanch:2002vj}
            to compute the $\pi\pi$ scattering amplitude.
            Our goal is to generalize that approach to accommodate arbitrary interaction kernels and,
            in particular, to describe scattering on baryons as well.

            The manuscript is organized as follows. In Sec.~\ref{sec:bses}
            we provide a brief overview of the covariant bound-state approach.
            The gauging method that provides the link between a hadron's current
            and the underlying quark-antiquark vertices is discussed in Sec.~\ref{sec:currents}
            and exemplified for the derivation of hadron form factors.
            In Sec.~\ref{sec:scattering} we derive the relation for scattering amplitudes,
            examine their ingredients in various kinematical limits and provide specific examples
            for scattering on baryons and mesons. We discuss possible applications and
            their practical feasibility in Sec.~\ref{sec:discussion} and conclude in Sec.~\ref{sec:conclusion}.

%            Taking left and right residues at the bound-state poles

        \section{Bound-state equations}\label{sec:bses}

            In order to collect the necessary tools for the derivation of the scattering amplitudes, we
            recall some basic relations of the covariant bound-state approach in this section.
            More detailed overviews can be found in Refs.~\cite{Loring:2001kv,Roberts:2007jh,Eichmann:2009zx,Blank:2011qk}.
            The following considerations are general and apply for any bound state of $n$ valence quarks and/or antiquarks.
            In practice, of course, we are primarily interested in mesons with $n=2$ and baryons with $n=3$.

            The quantity in QCD which describes a hadron microscopically
            is the $n-$quark (i.e., $2n-$point) Green function $G$, as well as
            its $n-$quark connected and amputated counterpart, the scattering matrix $T$ that is defined via
            \begin{equation}\label{T-definition}
                G = G_0 + G_0\,T\,G_0\,.
            \end{equation}
            Here, $G_0$ denotes the disconnected product of $n$ dressed quark propagators, e.g.: $G_0 = S \otimes S \otimes S$ in
            the case of a baryon.
            To keep the discussion as transparent as possible
            we will use a symbolic notation throughout the paper.
            All Dirac-Lorentz, color and flavor indices as well as momentum dependencies are suppressed,
            and the products in Eq.~\eqref{T-definition} and subsequent equations are understood as four-momentum integrations over all internal loop momenta.

       \begin{figure*}[t]
                    \begin{center}

                    \includegraphics[scale=0.12]{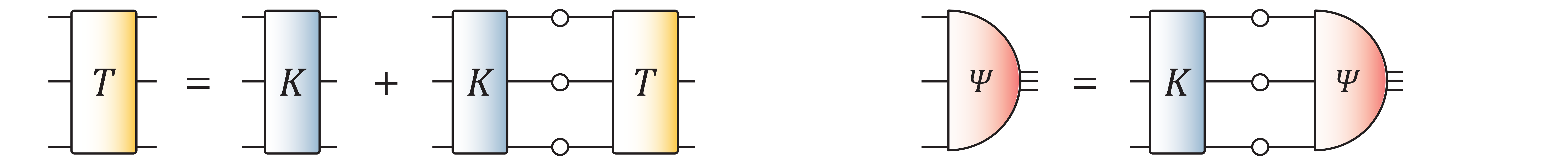}
                    \caption{(Color online) \textit{Left panel:} Dyson equation~\eqref{Dyson-for-T-1} for the scattering matrix in the three-quark system.
                                            \textit{Right panel:} bound-state equation~\eqref{boundstate-eq} for a baryon.   } \label{fig:bse}

                    \end{center}
        \end{figure*}

            The Green function $G$ can be expanded in its Dyson series, thereby defining the $n-$quark kernel $K$
            that is the sum of $m-$quark irreducible components ($2\leq m \leq n$):
            \begin{equation}
                G = G_0 + G_0\,K\,G_0 + G_0\,K\,G_0\,K\,G_0 + \dots\,.
            \end{equation}
            Its nonperturbative resummation yields Dyson's equation:
            \begin{align}
                & G = G_0 + G_0\,K\,G \quad \Leftrightarrow \label{Dyson-for-G-1}\\
                & G^{-1} = G_0^{-1} - K\,.
            \end{align}
            The equivalent series for the $T-$matrix reads
            \begin{equation}
                T = K + K\,G_0\,K + \dots\,,
            \end{equation}
            with its resummed version, illustrated in Fig.~\ref{fig:bse}:
            \begin{align}
                & T = K + K\,G_0\,T \quad \Leftrightarrow \label{Dyson-for-T-1}\\
                & T^{-1} = K^{-1} - G_0\,. \label{Dyson-for-T-2}
            \end{align}

            So far we have not gained much: we have related one unknown quantity ($G$, or $T$) with another ($K$).
            The central observation is that
            hadrons must appear as poles in the respective $n-$quark Green function or, equivalently,
            in the $n-$quark scattering matrix. These poles will be generated selfconsistently upon solving Eq.~\eqref{Dyson-for-T-1}, and they
            may be real and timelike or shifted into the timelike complex plane via a non-zero decay width.
             A pole in the scattering matrix defines a 'bound state' on its mass shell $P^2=-M^2$.
             The scattering matrix at the pole assumes the form
             \begin{equation}\label{poles-in-T}
                T\stackrel{P^2=-M^2}{\longlongrightarrow} \frac{\Psi\,\conjg{\Psi}}{P^2+M^2} + \text{regular terms}\,,
             \end{equation}
             where $\Psi$ defines the hadron's bound-state amplitude and $\conjg{\Psi}$ is its charge conjugate.
             The same relation holds for the Green function $G$ if the amplitude
             is replaced by the covariant wave function $G_0\Psi$, i.e., with quark propagator legs attached.
             At the pole, upon inserting Eq.~\eqref{poles-in-T} in Dyson's equation and comparing the residues of the singular terms,
             Eq.~\eqref{Dyson-for-T-1} reduces to a homogeneous bound-state equation for the amplitude $\Psi$:
             \begin{equation}\label{boundstate-eq}
                 K \,G_0\,\Psi = \Psi \quad \Leftrightarrow \quad T^{-1} \Psi = 0\,,
             \end{equation}
             where we have used Eq.~\eqref{Dyson-for-T-2} to arrive at the second version.
             Eq.~\eqref{boundstate-eq} is the homogeneous Bethe-Salpeter equation for a bound state of $n$ valence quarks and
             shown in Fig.~\ref{fig:bse} for the case of a baryon.
             It can be solved once the kernel $K$ and the dressed quark propagator $S$ that enters $G_0$ have been determined.

             The total hadron momentum $P^2$ enters Eq.~\eqref{boundstate-eq} as an external variable.
             The equation can be viewed as an eigenvalue problem for the quantity $K G_0$
             with eigenvalues $\lambda_i(P^2)$~\cite{Nakanishi:1969ph,Blank:2010bp}:
             \begin{align}
                & K \,G_0 \Psi_i = \lambda_i \Psi_i \quad \Leftrightarrow \\
                & T^{-1} \Psi_i = (\lambda_i^{-1}-1)\,G_0 \Psi_i\,. \label{boundstate-eq-lambda-2}
             \end{align}
             The eigenvector $\Psi_i$ describes a ground or excited-state hadron, with mass $M_i$, if its
             eigenvalue satisfies the condition $\lambda_i(P^2=-M_i^2)=1$.
             The ground state corresponds to the largest eigenvalue of the matrix $K G_0$.

             The bound-state equation determines a hadron amplitude up to a normalization.
             A covariant normalization follows from the requirement that the $T-$matrix has unit residue at the pole, i.e.
             \begin{equation}\label{T'}
                T\,'\stackrel{P^2=-M^2}{\longlongrightarrow} -\frac{\Psi\,\conjg{\Psi}}{(P^2+M^2)^2} + \text{less singular terms}\,,
             \end{equation}
             where $'$ denotes the derivative $d/dP^2$. Evaluating the relation
             \begin{equation}
                (T\,T^{-1})' = 0 \quad \Rightarrow \quad T' = -T \,(T^{-1})' \,T
             \end{equation}
             at the hadron pole yields, in combination with Eqs.~\eqref{poles-in-T} and \eqref{T'},
             the on-shell canonical normalization condition:
             \begin{equation}\label{canonical-norm}
                \conjg{\Psi} \left(T^{-1}\right)' \Psi = 1 \,.
             \end{equation}
             As a corollary of Eq.~\eqref{boundstate-eq-lambda-2},
             the derivative of the kernel $K$ that enters the normalization integral via Eq.~\eqref{Dyson-for-T-2}  can be traded
             for the derivative of the eigenvalue on the mass shell, so that the normalization simplifies to:
             $-\lambda'\,\conjg{\Psi} \,G_0 \Psi = 1$~\cite{Nakanishi:1965zza,Fischer:2009jm}.

             The numerical effort in solving a hadron's bound-state equation~\eqref{boundstate-eq}
             is related to the Poincar\'e-covariant structure of the hadron amplitudes.
             While a pion amplitude $\Psi_{\alpha\beta}(p,P)$ depends on four covariant basis elements,
             a nucleon amplitude $\Psi_{\alpha\beta\gamma\delta}(p,q,P)$ includes already 64~\cite{Eichmann:2009qa}.
             Here, Greek subscripts are Dirac indices and $p$, $q$ are relative four-momenta.
             In fact, the ability to solve the equations including the full covariant structure of the amplitudes is a prerequisite
             for computing form factors and scattering amplitudes.
             While the structure of ground-state baryons such as the nucleon and the $\Delta$-baryon is dominated by $s$ waves,
             orbital angular momentum in terms of $p$ waves, which are a consequence of Poincar\'e covariance,
             plays an important role as well~\cite{Oettel:1998bk,Alkofer:2004yf,Eichmann:2011vu,SanchisAlepuz:2011jn}.

       \begin{figure*}[t]
                    \begin{center}

                    \includegraphics[scale=0.12]{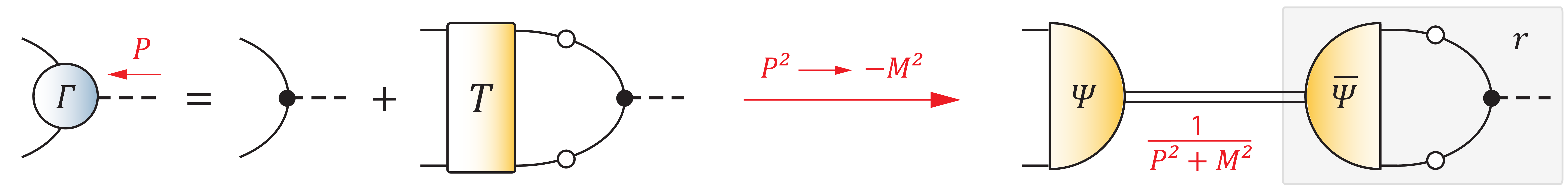}
                    \caption{(Color online) Quark-antiquark vertex~\eqref{vertex-definition} and its behavior
                             on the meson bound-state pole, Eq.~\eqref{poles-in-vertex}.   } \label{fig:vertex}

                    \end{center}
        \end{figure*}

             The physics that generates these phenomena is encoded in the dressed quark propagator $S$
             and the kernel $K$ which represent the input of the equations, cf.~Fig.~\ref{fig:bse}.
             The quark propagator and the one-particle-irreducible counterpart of $K$
             are subject to the Dyson-Schwinger equations and, in principle, can be solved numerically within a suitable truncation.
             Moreover, they are related by vector and axial-vector Ward-Takahashi identities
             which ensure electromagnetic current conservation as well as Gell-Mann-Oakes-Renner and Goldberger-Treiman relations at the hadron level.

             These identities can also be utilized as a systematic construction principle for kernel ans\"atze~\cite{Munczek:1994zz}.
             In practice, the majority of calculations in this particular framework have been performed in the rainbow-ladder truncation,
             where the $q\bar{q}$ kernel is given by a dressed gluon-ladder exchange, with two bare structures for the quark-gluon vertices
             and an effective gluon propagator built from the gluon dressing function and a simple model for the vertex dressing; see~\cite{Maris:2003vk,Maris:2005tt,Krassnigg:2009zh} for a selection of results.
             The quark propagator can then be easily solved from its DSE.
             Similarly, solutions of a baryon's bound-state equation, which involves the three-quark kernel
             \begin{equation}\label{three-body-kernel}
                 K = K_{(3)} + \big( S^{-1}\otimes K_{(2)} + \text{permutations}\big) \,,
             \end{equation}
             were obtained upon neglecting the $qqq-$irreducible part $K_{(3)}$,
             thereby defining the covariant Faddeev equation, and by using a rainbow-ladder ansatz for
             the $qq$ contribution $K_{(2)}$~\cite{Eichmann:2009qa,Eichmann:2011vu,SanchisAlepuz:2011jn}.
             Recent progress in constructing $q\bar{q}$ kernels beyond rainbow-ladder, without resorting to an order-by-order resummation,
             together with promising results in the meson sector~\cite{Alkofer:2008tt,Alkofer:2008et,Fischer:2008wy,Fischer:2009jm,Chang:2009zb,Chang:2010hb}
             provide hope that such expressions may eventually also be suitable for an implementation in baryon calculations.

             We wish to emphasize that the subsequent derivation is completely general and does not rely on any specific assumptions for the input $S$, $K_{(2)}$ and $K_{(3)}$ of the bound-state equations.
             We will treat these quantities essentially as black boxes and assume that they are known in advance.
             While the rainbow-ladder truncation will ultimately prove useful for practical implementations,
             the derivation of the scattering amplitudes is valid for general kernels.

        \section{Hadron currents}\label{sec:currents}

        \subsection{Quark-antiquark vertices}\label{sec:vertices}

            In order to relate the coupling of an external current,
            for instance, a pseudoscalar, vector or axialvector current,
            to the underlying description of the hadron as a composite object,
            one must specify how that current couples to the hadron's constituents.
            At the microscopic level, the current couples to quarks.
            That is described by the quark-antiquark vertex $\Gamma^\mu$, cf.~Fig.~\ref{fig:vertex}:
            \begin{equation}\label{vertex-definition}
                \Gamma^\mu := G_0^{-1}\,G\,\Gamma_0^\mu = \Gamma_0^\mu + T\,G_0 \,\Gamma_0^\mu\,,
            \end{equation}
            where in the second step we have exploited Eq.~\eqref{T-definition}.
            The quantities $G$, $G_0$, $T$ and $K$ are those introduced in Section~\ref{sec:bses} in the quark-antiquark case.
            The vertex is the contraction of the quark-antiquark Green function with a 'bare' term $\Gamma_0^\mu$, where
            $\mu$ is not necessarily a Lorentz index but rather a label for the type of current
            that will be identified with the 'gauging' index in Section~\ref{sec:gauging}.
            For instance, $\Gamma_0^\mu$ can represent $\gamma_5$, $\gamma^\mu$,
            or $\gamma_5\gamma^\mu$, with appropriate renormalization constants attached.

            Inserting Dyson's equation for $G$ in~\eqref{vertex-definition}
            yields the inhomogeneous Bethe-Salpeter equation for the vertex:
            \begin{equation}\label{ibse}
                \Gamma^\mu = \Gamma_0^\mu + K\,G_0\,\Gamma^\mu\,,
            \end{equation}
            which features the same ingredients as a meson's bound-state equation
            and thus can be solved consistently within the same truncation.

            Since the $q\bar{q}$ scattering matrix $T$ contains meson bound-state poles,
            with the pole behavior from Eq.~\eqref{poles-in-T},
            any pole will also appear in the vertex according to~\eqref{vertex-definition},
            so that on the mass shell of the total momentum the vertex becomes proportional to the bound-state amplitude:
            \begin{equation}\label{poles-in-vertex}
                \Gamma^\mu \stackrel{P^2=-M^2}{\longlongrightarrow}\frac{\Psi\,\conjg{\Psi}}{P^2+M^2}\,G_0 \,\Gamma_0^\mu =: \frac{r^\mu  \Psi}{P^2+M^2}\,.
            \end{equation}
            This relation holds as long as the respective meson wave function $\conjg{\Psi}\,G_0$
            has non-vanishing overlap with the structure $\Gamma_0^\mu$
            and the residue $r^\mu=\conjg{\Psi}\,G_0\,\Gamma_0^\mu$ is thus non-zero.
            Among other consequences, this feature is the underlying reason for 'vector-meson dominance' in hadron electromagnetic form factors:
            at the microscopic level, the coupling to a photon is represented by the quark-photon vertex which is a Lorentz vector.
            The bare term $\gamma^\mu$ has an overlap with the $\rho-$meson wave function, and that overlap is proportional to its electroweak decay constant.
            Consequently, the $\rho-$meson pole appears in the (transverse part of the) quark-photon vertex and transpires to the level of hadron form factors.
            If the quark-antiquark kernel $K$ that enters Eqs.~\eqref{boundstate-eq} and~\eqref{ibse} as an input allows for a $\rho\rightarrow\pi\pi$ decay,
            the $\rho-$meson acquires a width and its pole
            will be shifted into the complex plane.

            Eq.~\eqref{poles-in-vertex} further implies that a hadron's coupling to a meson
            can be described by the respective quark-antiquark vertex
            in the same manner as the interaction with a photon.
            Upon removing the onshell meson pole together with its residue from the vertex,
            the coupling to the quark is represented by the meson's bound-state amplitude.
            This is the central feature that enables a common description of hadron-meson and hadron-photon interactions:
            with a single expression for the $q\bar{q}$ kernel $K$, different types of quark-antiquark vertices
            can be readily solved from the same inhomogeneous BSE~\eqref{ibse}
            which differs only by the inhomogeneous driving term $\Gamma_0^\mu$.
            In view of investigating the hadron's coupling to a photon,
            one solves the BSE with the inhomogeneous term $\gamma^\mu$ to obtain the quark-photon vertex;
            for describing its coupling to a pion, one would implement $\gamma_5$ and remove the pion pole
            together with its residue from the resulting pseudoscalar vertex,
            or solve the pion's bound-state equation~\eqref{boundstate-eq} directly.
            Apart from that, there is no conceptual difference in the description.
            Thus, whenever a quark-antiquark vertex appears in the following considerations,
            we will leave its Lorentz type unspecified.

        \subsection{Gauging of equations}\label{sec:gauging}

            While, at least in principle, the tower of QCD's Green functions can be solved from Dyson-Schwinger equations,
            the question of how to combine these Green functions for obtaining structure functions at the hadron level
            is already much less straightforward to answer.
            A systematic construction principle to derive the coupling of a hadron to an external current
            is the 'gauging of equations' method~\cite{Haberzettl:1997jg,Kvinikhidze:1998xn,Kvinikhidze:1999xp}.

             Diagrammatically, gauging corresponds to the coupling of an external current with given quantum numbers to an $n-$point Green function.
             When acting upon that Green function it yields an $(n+1)-$point function with an additional leg.
             Gauging is formally indicated by an index $\mu$ and has the properties of a derivative, i.e. it is linear and satisfies Leibniz' rule.
             At the quark level, it amounts to replacing each inverse dressed quark propagator by the respective quark-antiquark vertex,
             i.e., the vertex $\Gamma^\mu$ is the gauged inverse quark propagator:
             \begin{equation}
                 S^{-1} \quad \longrightarrow \quad \left(S^{-1}\right)^\mu := \Gamma^\mu = (1+T\,G_0)\,\Gamma_0^\mu\,.
             \end{equation}
             Correspondingly, each quark propagator is replaced by $S^\mu = -S\,\Gamma^\mu S$, where
             the specific type of gauging is determined by the structure $\Gamma_0^\mu$.

             To derive a quark-level description of a hadron's coupling to an external current (or several external currents),
             one applies the gauging operator to the $n-$quark scattering matrix $T$. In the same way as the pole residues in $T$
             correspond to hadron bound-state amplitudes, the residue of $T^\mu$ defines a hadron's current that,
             depending on the type of $\Gamma_0^\mu$, includes the various form factors of a hadron:
             for example, its electromagnetic ($\gamma^\mu$), pseudoscalar ($\gamma_5$) or axial form factors ($\gamma_5 \gamma^\mu$).
             If $T$ is gauged twice, the residue in $T^{\mu\nu}$ is the desired scattering amplitude that describes,
             for instance, Compton scattering or the scattering of a nucleon and a pion.
             With the help of Dyson's equation~\eqref{Dyson-for-T-1} and the Leibniz rule, the gauging operation can be systematically
             traced back to the gauging of quarks and that of the kernel $K$.

             Before applying the formalism to determine scattering amplitudes,
             we will revisit the derivation of hadron form factors in the next section.
             We will make frequent use of the derivative property of the gauging operator which implies
             \begin{equation}\label{T-Tinv}
                  T^\mu   =  -T\left(T^{-1}\right)^\mu T
             \end{equation}
             and analogous relations for other Green functions.

        \subsection{Derivation of form factors}\label{sec:ffs}

            The form factors of a hadron are the Lorentz-invariant dressing functions
            of the hadron's current matrix element $J^\mu$.
            The latter is a three-point function
            that describes the hadron's interaction with the external current.
            At the level of QCD's Green functions, that interaction
            is contained in the gauged $n-$quark scattering matrix, i.e., in the $(2n+1)-$point function $T^\mu$.
            To isolate a hadron's form factor inside $T^\mu$, one must inspect its pole structure that is given in terms of
             a pole for the incoming and one pole for the outgoing hadron.
             The pole residue defines the current matrix element:
             \begin{equation}\label{poles-in-Tmu}
                 T^\mu \stackrel{P_i^2=P_f^2=-M^2}{\longlonglonglongrightarrow} - \frac{\Psi_f\,J^\mu \,\conjg{\Psi}_i}{(P_f^2+M^2)(P_i^2+M^2)}\,,
             \end{equation}
             where $(P_f-P_i)^2=Q^2$ is the four-momentum transfer and,
             for instance in the baryon case, $\Psi_i = \Psi(p_i,q_i,P_i)$ and
             $\Psi_f = \Psi(p_f,q_f,P_f)$ are in- and outgoing baryon amplitudes with different
             momentum dependencies.
             On the other hand, Eqs.~\eqref{T-Tinv} and~\eqref{poles-in-T} yield
             \begin{equation}
             \begin{split}
                  T^\mu   = & -T\left(T^{-1}\right)^\mu T \stackrel{P_i^2=P_f^2=-M^2}{\longlonglonglongrightarrow} \\
                       &    -\frac{\Psi_f\,\conjg{\Psi}_f}{P_f^2+M^2}\left(T^{-1}\right)^\mu\frac{\Psi_i\,\conjg{\Psi}_i}{P_i^2+M^2}\,,
             \end{split}
             \end{equation}
             and by comparing these two equations one obtains the current as
             the gauged inverse scattering matrix element between the onshell hadron amplitudes:
             \begin{equation}\label{emcurrent-gauging}
                J^\mu = \conjg{\Psi}_f \left(T^{-1}\right)^\mu \Psi_i \,.
             \end{equation}
             In the limit $Q^2\rightarrow 0$, Eq.~\eqref{emcurrent-gauging} reproduces the canonical normalization condition:
             once the bound-state amplitude is normalized via Eq.~\eqref{canonical-norm},
             the normalization of the resulting form factors is fixed as well.
             If the vector WTI is respected throughout every stage,
             electromagnetic current conservation and thus charge conservation, e.g., for a pion or a proton, readily follow~\cite{Oettel:1999gc}.
             Similarly, if the axialvector WTI is correctly implemented at the level of the kernels,
             the axial form factors satisfy Goldberger-Treiman relations~\cite{EF-Axial}.

             To make Eq.~\eqref{emcurrent-gauging} more useful for practical numerical implementations,
             we need to express the inverse scattering matrix by the kernel $K$ and the propagator product $G_0$.
             Gauging Eq.~\eqref{Dyson-for-T-2} yields
             \begin{equation}\label{T-1mu}
             \begin{split}
                 \left(T^{-1}\right)^\mu &= \left(K^{-1}\right)^\mu - G_0^\mu  \\
                                         &= G_0 \,\mathbf{\Gamma}^\mu\, G_0 -K^{-1} K^\mu K^{-1} \,,
             \end{split}
             \end{equation}
             where we have denoted the gauged inverse propagator product by $\left(G_0^{-1}\right)^\mu=:\mathbf{\Gamma}^\mu$.
             For instance, in a two-body system it is given by
             \begin{equation}\label{Gamma-mu-decomposition-2body}
                 \mathbf{\Gamma}^\mu = \left( S^{-1} \otimes S^{-1} \right)^\mu = \Gamma^\mu \otimes S^{-1} + S^{-1} \otimes \Gamma^\mu\,,
             \end{equation}
             and thus
             \begin{equation}
                 G_0\,\mathbf{\Gamma}^\mu \,G_0 = S\,\Gamma^\mu S \otimes S + S \otimes S\,\Gamma^\mu S\,.
             \end{equation}

             When acting on the onshell amplitudes, the inverse kernels in~\eqref{T-1mu} can be replaced
             by $G_0$ via the bound-state equations $K^{-1} \Psi_i = G_0 \Psi_i$ and
             $\conjg{\Psi}_f K^{-1} = \conjg{\Psi}_f \,G_0$. This yields the final result for the current:
             \begin{equation}\label{current}
                J^\mu = \conjg{\Psi}_f \,G_0 \left(\mathbf{\Gamma}^\mu - K^\mu \right) G_0 \,\Psi_i \,.
             \end{equation}
             This relation has been exploited in numerous meson form-factor studies~\cite{Maris:1999bh,Bhagwat:2006pu,Jarecke:2002xd,Mader:2011zf}.
             In rainbow-ladder truncation: $K_{(2)}^\mu=0$, and the resulting meson current is expressed by
             impulse-approximation diagrams alone. In the baryon case, the three-body kernel of Eq.~\eqref{three-body-kernel} with $K_{(3)}=0$
             yields the covariant Faddeev equation. The gauged kernel in rainbow-ladder becomes
             \begin{equation}
                 K^\mu = \Gamma^\mu \otimes K_{(2)} \quad + \text{permutations}\,,
             \end{equation}
             which was recently implemented to compute the nucleon's electromagnetic form factors~\cite{Eichmann:2011vu}.  % and axial
             The analogue of Eq.~\eqref{current} in the quark-diquark model was utilized in various
             quark-diquark studies of baryon form factors~\cite{Oettel:1999gc,Oettel:2000jj,Cloet:2008re,Eichmann:2009zx}.

       \begin{figure*}[t]
                    \begin{center}

                    \includegraphics[scale=0.10]{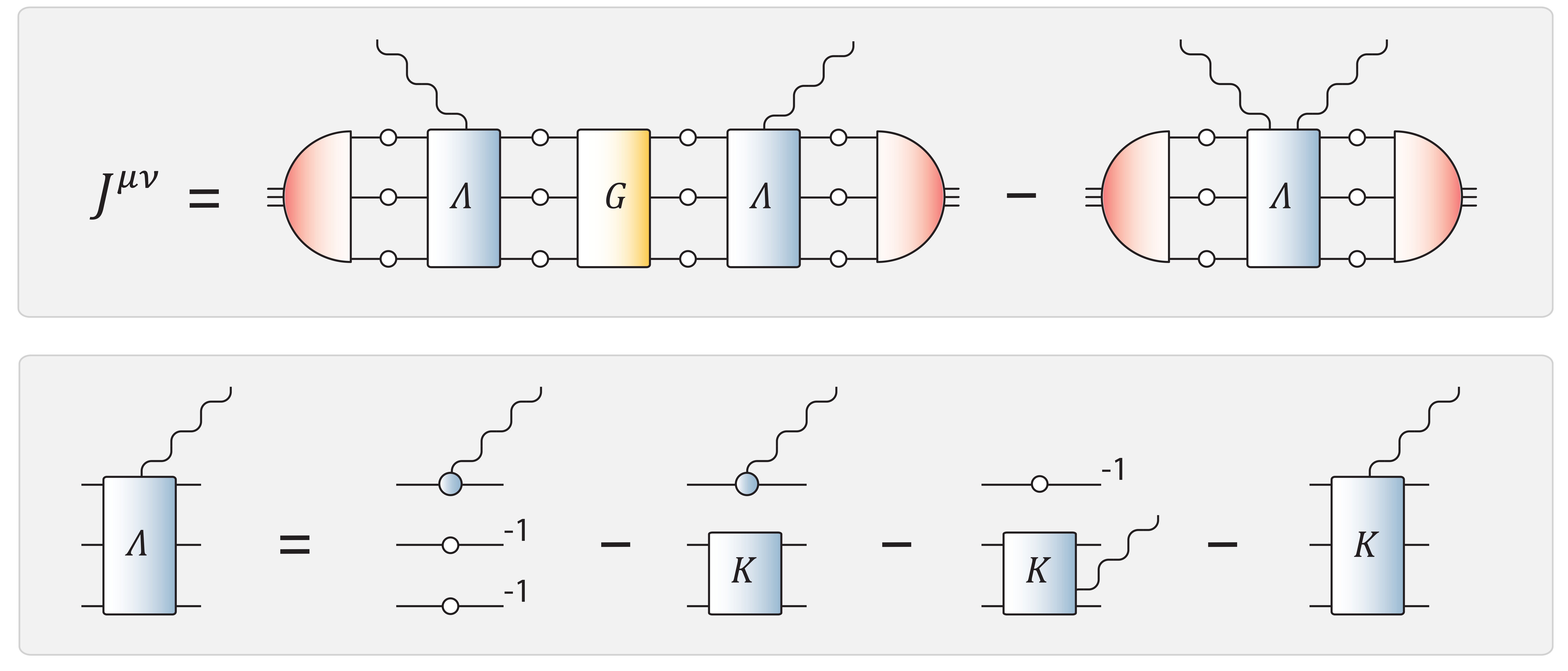}
                    \caption{(Color online) \textit{Upper panel:} Diagrammatic representation of the scattering amplitude $J^{\mu\nu}$,
                                            Eq.~\eqref{scattering-final}, in the case of a baryon.
                                            A symmetrization of the two currents with indices $\mu$ and $\nu$ leads to an additional crossed diagram.
                                            \textit{Lower panel:} Decomposition of the quantity $\mathbf{\Lambda}^\mu$,
                                            cf. Eqs.~\eqref{lambda}, \eqref{Gamma-mu-3body} and \eqref{K-mu-3body}, that enters the scattering amplitude.
                                            For each of the first three graphs on the r.h.s. there are two further permutations with respect to the quark lines.
                                            Only the first two diagrams contribute in rainbow-ladder truncation.    } \label{fig:current+scattering}

                    \end{center}
        \end{figure*}

        \section{Hadron scattering}\label{sec:scattering}

        \subsection{Derivation of the scattering amplitude}\label{sec:scattering-derivation}

             We now turn to the derivation of four-point functions within the present formalism.
             Generalizing Eq.~\eqref{poles-in-Tmu}, one can identify the hadron's coupling to two external currents as the
             residue of the scattering matrix $T^{\mu\nu}$ that is gauged \textit{twice}:
             \begin{equation}
                 T^{\mu\nu} \stackrel{P_i^2=P_f^2=-M^2}{\longlonglonglongrightarrow}  \frac{\Psi_f\,J^{\mu\nu} \,\conjg{\Psi}_i}{(P_f^2+M^2)(P_i^2+M^2)}\,.
             \end{equation}
             $J^{\mu\nu}$ is the desired scattering matrix on the mass shell of the incoming and outgoing hadron.
             The strategy to obtain its quark-level decomposition proceeds along the same lines as for the
             form factors in the previous section.
             After expressing $J^{\mu\nu}$ as a function of $T$, $T^\mu$ and $T^{\mu\nu}$,
             we will relate these quantities to the dressed quark propagator $S$ and the kernel $K$
             as well as their gauged analogues.

             In the first step we use the derivative property~\eqref{T-Tinv} of the gauging operation to rewrite $T^{\mu\nu}$:
             \begin{equation}
             \begin{split}
                 T^{\mu\nu} =& \left[-T\left(T^{-1}\right)^\mu T\right]^\nu \\
                            =& - T\left(T^{-1}\right)^{\mu\nu} T  \\
                             & - T\left(T^{-1}\right)^\mu T^\nu - T^\nu \left(T^{-1}\right)^\mu T \\
                            =& - T \left(T^{-1}\right)^{\mu\nu} T \\
                             & + T \left(T^{-1}\right)^\mu T \left(T^{-1}\right)^\nu T \\
                             & + T \left(T^{-1}\right)^\nu T \left(T^{-1}\right)^\mu T \\
                            =& \; T \left[  \left(T^{-1}\right)^{\{\mu} T \left(T^{-1}\right)^{\nu\}} - \left(T^{-1}\right)^{\mu\nu} \right] T\,,
             \end{split}
             \end{equation}
             where the curly brackets denote a symmetrization in the indices $\mu$ and $\nu$.
             Upon replacing the $T-$matrices on the far left and right with
             \begin{equation}
                 \frac{ \Psi_f \conjg{\Psi}_f}{P_f^2+M^2}\quad \text{and} \quad
                 \frac{ \Psi_i \conjg{\Psi}_i}{P_i^2+M^2}\,,
             \end{equation}
             respectively, we can identify the onshell residue $J^{\mu\nu}$:
             \begin{equation}\label{scattering-gauging}
                 J^{\mu\nu} = \conjg{\Psi}_f \left[ \left(T^{-1}\right)^{\{\mu} T \left(T^{-1}\right)^{\nu\}} - \left(T^{-1}\right)^{\mu\nu} \right] \Psi_i \,.
             \end{equation}
             This equation is the equivalent of Eq.~\eqref{emcurrent-gauging} for a four-point function.

             In the second step we want to relate the scattering amplitude $J^{\mu\nu}$ to the quantities $K$ and $G_0$.
             To obtain the first term in Eq.~\eqref{scattering-gauging}, we insert $(T^{-1})^\mu$ from Eq.~\eqref{T-1mu} and use the relations
             \begin{equation}
             \begin{split}
                  & G_0 \,T \,G_0 =  G - G_0\,, \\
                  & K^{-1} T \,G_0 = G_0 \,T K^{-1} = G\,,\\
                  & K^{-1} T K^{-1} = G + K^{-1}\,,
             \end{split}
             \end{equation}
             that follow from the definition of $T$ in \eqref{T-definition} and Dyson's equation~\eqref{Dyson-for-G-1}.
             Note that the arrangement of the terms in the resummation is
             irrelevant: $G_0 \,K \,G = G \,K \,G_0$. Again, the inverse kernels can be replaced
             by $G_0$ when acting on the bound-state amplitudes. The result is
             \begin{equation}\label{T-1-munu-1}
             \begin{split}
                  &\left(T^{-1}\right)^{\mu} T \left(T^{-1}\right)^{\nu} =  G_0 \Big[  \mathbf{\Gamma}^\mu (G-G_0) \,\mathbf{\Gamma}^\nu  \\
                  &+ K^\mu (G+K^{-1})\, K^\nu   - K^\mu  G \,\mathbf{\Gamma}^\nu - \mathbf{\Gamma}^\mu  G \,K^\nu \Big] \,G_0\,.
             \end{split}
             \end{equation}
             For the second term in~\eqref{scattering-gauging}, we gauge $(T^{-1})^\mu$ from Eq.~\eqref{T-1mu} once again:
             \begin{equation}\label{T-1-munu-2}
             \begin{split}
                 \left(T^{-1}\right)^{\mu\nu}  &= G_0 \big[ \mathbf{\Gamma}^{\mu\nu} - \mathbf{\Gamma}^{\{\mu} \,G_0 \,\mathbf{\Gamma}^{\nu\}} \big] G_0 \\
                                               & + K^{-1} \big[ K^{\{\mu} K^{-1} K^{\nu\}}  - K^{\mu\nu}\big] K^{-1} \,.
             \end{split}
             \end{equation}
             Some of the terms cancel when summing up Eqs.~\eqref{T-1-munu-1} and~\eqref{T-1-munu-2}, and the remainder is
             \begin{equation}\label{scattering-gauging-2}
             \begin{split}
                 J^{\mu\nu} =  \conjg{\Psi}_f \,G_0 \,\Big[ & (\mathbf{\Gamma}-K)^{\{\mu} \, G \,(\mathbf{\Gamma}-K)^{\nu\}}  \\[-2mm]
                 &  -(\mathbf{\Gamma}^{\mu\nu} - K^{\mu\nu})  \Big] \,G_0 \Psi_i \,.
             \end{split}
             \end{equation}
             With the shorthand notation
             \begin{equation} \label{lambda}
                 \mathbf{\Lambda}^\mu:= \mathbf{\Gamma}^\mu - K^\mu\,, \quad
                 \mathbf{\Lambda}^{\mu\nu}= \mathbf{\Gamma}^{\mu\nu} - K^{\mu\nu}\,,
             \end{equation}
             the final result for the current~\eqref{current} and the scattering amplitude~\eqref{scattering-gauging-2}
             can be written in the compact form
             \begin{align}
                 J^\mu &= \conjg{\Psi}_f \,G_0 \,\mathbf{\Lambda}^\mu \,G_0 \Psi_i\,, \label{current-final}\\
                 J^{\mu\nu} &= \conjg{\Psi}_f \,G_0 \left[  \mathbf{\Lambda}^{\{\mu} G \,\mathbf{\Lambda}^{\nu\}} - \mathbf{\Lambda}^{\mu\nu} \right] G_0 \Psi_i\,. \label{scattering-final}
             \end{align}
             Eq.~\eqref{scattering-final} is the central result of this work.
             We reiterate that the equation is completely general: we have not assumed any specific form
             for the type of currents, or the type of hadron, or the structure of the kernel that enters the equation.
             Fig.~\ref{fig:current+scattering} illustrates the relation for the case of a baryon.
             We will examine its physical content in the next section.

        \subsection{Ingredients of $J^\mu$ and $J^{\mu\nu}$} \label{sec:ingredients}

             To analyze the features of the scattering amplitude $J^{\mu\nu}$ in Eq.~\eqref{scattering-final},
             we need to examine its ingredients in more detail. It involves the quantities
             \begin{equation}\label{scattering-ingredients}
                 \mathbf{\Gamma}^\mu\,, \quad
                 \mathbf{\Gamma}^{\mu\nu}\,, \quad
                 K^\mu\,, \quad
                 K^{\mu\nu}\,, \quad
                 G\,,
             \end{equation}
             i.e., the gauged disconnected propagator products, the gauged kernels,
             and the $n-$quark Green function $G$.
             We have assumed knowledge of the kernel $K$, given in terms of a diagrammatic expression or an ansatz,
             and consequently of the quark propagator $S$, the quark-antiquark vertex $\Gamma^\mu$, and the bound-state amplitude $\Psi$.
             The propagator can be solved from its DSE which involves a quark-gluon vertex that is compatible with the expression for $K$;
             the vertex is solved from its inhomogeneous BSE~\eqref{ibse}; and the bound-state amplitudes are determined from their
             covariant bound-state equations.
             Thus, the remaining task is to express the quantities in Eq.~\eqref{scattering-ingredients} in terms of $S$, $\Gamma^\mu$ and $K$.

             We have already stated the decomposition of $\mathbf{\Gamma}^\mu$ for the quark-antiquark case in Eq.~\eqref{Gamma-mu-decomposition-2body};
             the generalization to the three-quark system is straightforward:
             \begin{equation}\label{Gamma-mu-3body}
                 \mathbf{\Gamma}^\mu  = \Gamma^\mu \otimes S^{-1}\otimes S^{-1} \quad + \;\text{2 permutations}\,.
             \end{equation}
             Let us now have a second look at Eq.~\eqref{scattering-final}.
             Due to the decomposition of $G$ from Eq.~\eqref{T-definition}
             we note that the first term contains a sum of a disconnected
             contribution $G_0$ and another part that involves the scattering
             matrix $T$. In combination with the decomposition of
             $\mathbf{\Gamma}^\mu$ contained in $\mathbf{\Lambda}^{\mu}$ via
             Eq.~\eqref{lambda}, the part with $G_0$ leads to diagrams
             in the baryon's scattering amplitude where the two currents
             couple to the same quark (handbag diagrams) or different quark
             lines (cat's-ears diagrams). On the other hand, the appearance
             of the scattering matrix $T$ in the other part holds the promise
             for recovering baryon resonances in the $s$ and $u$ channels.
             We will return to this observation in more detail in
             Section~\ref{sec:resonances}.

             Applying the gauging operation to the previous relation for $\mathbf{\Gamma}^\mu$ yields
             the quantity $\mathbf{\Gamma}^{\mu\nu}$ that enters the second term in~\eqref{scattering-final}. One obtains
             \begin{equation}
                 \mathbf{\Gamma}^{\mu\nu} = \Gamma^{\mu\nu}\otimes S^{-1} + S^{-1} \otimes \Gamma^{\mu\nu} + \Gamma^{\{\mu} \otimes \Gamma^{\nu\}}
             \end{equation}
             in the quark-antiquark case and
             \begin{equation}\label{Gamma-mu-nu-3body}
             \begin{aligned}
                 \mathbf{\Gamma}^{\mu\nu} & = \Gamma^{\mu\nu}\otimes S^{-1}\otimes S^{-1}   &+ \;\text{2 permutations} \\
                                   & + \Gamma^{\{\mu}\otimes \Gamma^{\nu\}} \otimes S^{-1}   &+ \;\text{2 permutations}
             \end{aligned}
             \end{equation}
             for a system of three quarks.
             Here we have introduced the vertex $\Gamma^{\mu\nu}$ that describes the coupling of a dressed quark
             to two external currents. We will analyze its features in detail in Section~\ref{sec:Gamma-mu-nu} and show that it is the source of
             those diagrams in $J^{\mu\nu}$ that, on a hadronic level, amount to $t-$channel meson exchange.
             The contributions from the second line of Eq.~\eqref{Gamma-mu-nu-3body} will
             provide further cat's-ears diagrams in the four-point function $J^{\mu\nu}$.

             The remaining quantities to be determined are the gauged kernels $K^\mu$ and $K^{\mu\nu}$. For a meson the respective construction is,
             at least in principle, straightforward: for a given diagrammatic expression of the quark-antiquark kernel $K_{(2)}$, the
             gauged kernel is obtained by replacing all internal dressed quark propagators by $-S\,\Gamma^\mu S$,
             and each further Green function by its gauged counterpart (as long as that object exists).
             The kernel in the baryon case, on the other hand, is the sum of two- and three-quark irreducible contributions given in Eq.~\eqref{three-body-kernel}.
             Therefore, the respective gauged kernel also picks up contributions proportional to $\Gamma^\mu$ and $\Gamma^{\mu\nu}$:
             \begin{alignat}{4}
                 K^\mu &= K_{(3)}^\mu &&+ \big( \Gamma^\mu \otimes K_{(2)} &&+ \text{perm.}\big) \nonumber \\
                       &   &&+ \big( S^{-1} \otimes K_{(2)}^\mu &&+ \text{perm.}\big)\,, \label{K-mu-3body}\\[2mm]
                 K^{\mu\nu} &=K_{(3)}^{\mu\nu} &&+ \big( \Gamma^{\mu\nu} \otimes K_{(2)} &&+ \text{perm.} \big) \nonumber\\
                            & &&+ \big( \Gamma^{\{\mu} \otimes K_{(2)}^{\nu\}} &&+ \text{perm.}\big) \nonumber\\
                            & &&+ \big( S^{-1}\otimes K_{(2)}^{\mu\nu} &&+ \text{perm.}\big)\,. \label{K-mu-nu-3body}
             \end{alignat}
             The object $\mathbf{\Lambda}^{\mu} = \mathbf{\Gamma}^{\mu} - K^\mu$, composed of
             Eqs.~\eqref{Gamma-mu-3body} and~\eqref{K-mu-3body}, is depicted in the lower panel of Fig.~\ref{fig:current+scattering}.

             \subsection{Hadron resonances}\label{sec:resonances}

             As already anticipated above, the first term in the scattering amplitude of Eq.~\eqref{scattering-final} exhibits a welcome feature:
             it involves the $n-$quark Green function $G$ which contains intermediate hadron resonances.
             For instance, if the two currents with labels $\mu$ and $\nu$ represent photons,
             $J^{\mu\nu}$ describes Compton scattering on a nucleon; if one of the two currents
             is of pseudoscalar nature, it describes pion photoproduction,
             and if both indices are pseudoscalar, one obtains the nucleon-pion scattering amplitude.
             In these cases, the Green function $G$ involves all possible $s-$channel nucleon resonances and,
             via symmetrization in the indices $\mu$ and $\nu$, also those in the $u$ channel.
             Comparison with Eqs.~\eqref{poles-in-T} and~\eqref{current-final}
             entails that at the respective pole locations $P^2=-M^2$ of the
             resonances the scattering amplitude $J^{\mu\nu}$ becomes
             the product of two currents:
             \begin{equation}\label{s-channel-poles}
             \begin{split}
                J^{\mu\nu} \stackrel{P^2=-M^2}{\longlongrightarrow} & \quad \conjg{\Psi}_f \,G_0 \, \mathbf{\Lambda}^{\{\mu} \frac{G_0\,\Psi\,\conjg{\Psi}\,G_0}{P^2+M^2} \,\mathbf{\Lambda}^{\nu\}} \,G_0 \Psi_i = \\
                                                                    & = \frac{J_f^{\{\mu}J_i^{\nu\}}}{P^2+M^2}\,,
             \end{split}
             \end{equation}
             where, depending on the order of the indices $\mu$ and $\nu$, $-P^2$ is equal to the Mandelstam variable $s$ or $u$.
             This is exactly the type of diagram illustrated in Fig.~\ref{fig:preliminary}c.

             Eq.~\eqref{scattering-final} additionally implements all offshell contributions of nucleon resonances through the three-quark Green function $G$.
             In practice, however, a fully self-consistent determination of $G$ through Dyson's equation~\eqref{Dyson-for-G-1} is not feasible
             due to its nature of being a six-point function.
             Owing to the rapidly increasing number of basis elements and independent momentum variables,
             present computational resources allow for a self-consistent computation of four-point functions at best,
             such as for example the nucleon bound-state amplitude,
             without truncating the covariant structure of the desired quantity.

             A potential alternative is to exploit Dyson's equation and derive an inhomogeneous Bethe-Salpeter equation for the five-point function
             \begin{equation}\label{resonances-bse-1}
                 \Psi_i^\nu :=G_0^{-1}\,G\,\mathbf{\Lambda}^\nu G_0 \,\Psi_i
             \end{equation}
             that enters the expression for $J^{\mu\nu}$.
             Using $G=G_0+G_0\,K\,G$, one arrives at the equation
             \begin{equation}\label{Psi-5point-eq}
                 \Psi_i^\nu = \mathbf{\Lambda}^\nu G_0 \,\Psi_i + K\,G_0\,\Psi_i^\nu
             \end{equation}
             which determines $\Psi_i^\nu$ self-consistently once the driving term $\mathbf{\Lambda}^\nu G_0 \,\Psi_i$ is known.
             The first term in the scattering amplitude~\eqref{scattering-final} then acquires the form
             \begin{equation}
                  \conjg{\Psi}_f \,G_0  \,\mathbf{\Lambda}^{\{\mu} \,G_0 \,\Psi_i^{\nu\}}  \,,
             \end{equation}
             and substituting the analogue of Eq.~\eqref{Psi-5point-eq} for $\conjg{\Psi}_f^\mu$ yields
             \begin{equation}\label{resonances-bse-4}
                  \conjg{\Psi}_f^{\{\mu} \,G_0  \,\Psi_i^{\nu\}} - \conjg{\Psi}_f^{\{\mu} \,G_0\,K\,G_0  \,\Psi_i^{\nu\}}  \,,
             \end{equation}
             where the interactions with the currents were completely absorbed in the five-point functions $\Psi_{i,f}^\mu$.

             Nevertheless, the attempt to solve Eq.~\eqref{Psi-5point-eq} still presents a numerical challenge.
             As a first step in exploring the features of Eq.~\eqref{scattering-final},
             a separable pole approximation in the spirit of Eq.~\eqref{s-channel-poles},
             with intermediate nucleon and $\Delta$ resonances, seems far more practical.
             In that respect it will be advantageous to separate the 'pure' pole contribution encapsulated in $T$ from the remaining non-resonant structure.
             Naturally, following such a course provides limited insight since one can only recover the effects of those resonances
             that were explicitly inserted into the diagram in the beginning.
             In view of the availability of $NN\gamma$, $NN\pi$, $N\Delta\gamma$ and $N\Delta\pi$ form factors
             in the rainbow-ladder truncated Dyson-Schwinger approach~\cite{Eichmann:2011vu,EF-Axial,EN-Delta,Mader:2011zf},
             however, such an attempt would still be worthwhile as a first step.
             On a more general note, even once Eq.~\eqref{Psi-5point-eq} is solved fully selfconsistently,
             the fidelity of the rainbow-ladder truncation in view of nucleon resonances beyond the $\Delta(1232)$ is essentially unknown.
             Thus, a self-consistent determination of $J^{\mu\nu}$ should also be augmented by an effort to go beyond rainbow-ladder.

       \begin{figure*}[t]
                    \begin{center}

                    \includegraphics[scale=0.10]{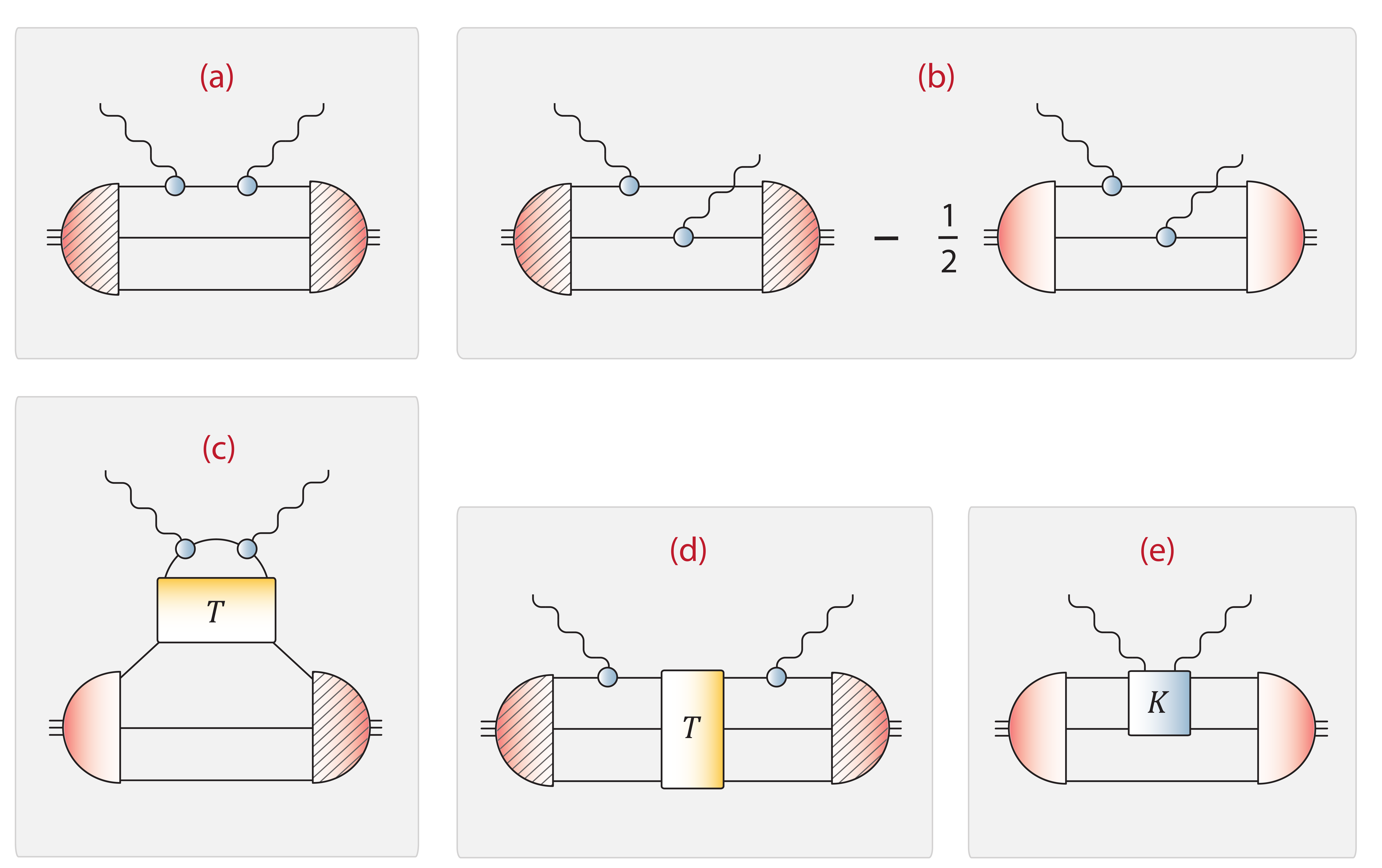}
                    \caption{(Color online) Contributions to the nucleon-photon Compton amplitude $J^{\mu\nu}$ in rainbow-ladder truncation,
                                            cf.~Eq.~\eqref{J-mu-nu-RL}. All quark propagators are dressed, and
                                            each box includes further diagrams with permuted quark lines.
                                            In addition, the photon indices $\mu$ and $\nu$ must be symmetrized.
                                            $T$ represents the quark-antiquark and three-quark scattering matrices and $K$ the two-quark irreducible kernel.
                                            The hatched nucleon amplitudes represent the quantities defined in~\eqref{hatched}.
                                            Diagram (c) containts $t-$channel meson exchange whereas $s-$ and $u-$channel nucleon resonances appear in diagram (d).
                                            To obtain the pion photoproduction amplitude, replace one of the quark-photon vertices by an onshell pion amplitude;
                                            for nucleon-pion scattering, replace both.
                                             } \label{fig:4-point-function}

                    \end{center}
        \end{figure*}

             \subsection{$\Gamma^{\mu\nu}$ and meson form factors}\label{sec:Gamma-mu-nu}

             In contrast to the kinematical structure of the scattering amplitude in the $s$ channel,
             the second contribution to Eq.~\eqref{scattering-final} which involves
             the four-point function $\Gamma^{\mu\nu}$ is easier to access. $\Gamma^{\mu\nu}$ describes the coupling of a dressed quark
             to two external currents.
             To derive an expression for that vertex, we start from the inhomogeneous BSE~\eqref{ibse} for the quark-antiquark vertex $\Gamma^\mu$.
             Gauging that equation with an additional index $\nu$ removes the constant inhomogeneous term and yields:
             \begin{equation}\label{ibse-Gamma-mu-nu}
             \begin{split}
                 \Gamma^{\mu\nu} &= (\Gamma^\mu)^\nu = ( \Gamma_0^\mu + K\,G_0 \,\Gamma^\mu)^\nu = (K\,G_0 \,\Gamma^\mu)^\nu \\
                                 &= \underbrace{K^\nu G_0\,\Gamma^\mu + K\,G_0^\nu \,\Gamma^\mu}_{\Gamma_0^{\mu\nu}} + K\,G_0\,\Gamma^{\mu\nu}\,.
             \end{split}
             \end{equation}
             Since the sequence of gauging is irrelevant, the quantities in the second line must be symmetric under exchange of $\mu$ and $\nu$.
             Eq.~\eqref{ibse-Gamma-mu-nu} is again an inhomogeneous Bethe-Salpeter equation, now for the quantity $\Gamma^{\mu\nu}$, with an inhomogeneous term
             \begin{equation}
                 \Gamma_0^{\mu\nu}:= K^\nu G_0\,\Gamma^\mu + K\,G_0^\nu \,\Gamma^\mu\,,
             \end{equation}
             and can be solved selfconsistently once the quark propagator, $\Gamma^\mu$ and $K^\mu$ are known.
             In comparison with the considerations of the previous subsection, 
             this equation is relatively easy to handle since $\Gamma^{\mu\nu}$ is merely a four-point function.

             Due to its structure of an inhomogeneous BSE, one can rearrange Eq.~\eqref{ibse-Gamma-mu-nu}
             to obtain a form analogous to~\eqref{vertex-definition}:
             \begin{equation}
                 \Gamma^{\mu\nu} = G_0^{-1} G\,\Gamma_0^{\mu\nu} = ( 1 + T\,G_0 )\,\Gamma_0^{\mu\nu}\,.
             \end{equation}
             From Eq.~\eqref{Dyson-for-T-1} one has $(1+T\,G_0)\,K = T$ and thus
             \begin{equation}\label{Gamma-mu-nu-T}
             \begin{split}
                 \Gamma^{\mu\nu} &= ( 1 + T\,G_0 )\,K^\nu G_0\,\Gamma^\mu + T\,G_0^\nu \,\Gamma^\mu = \\
                                 &= K^\nu G_0\,\Gamma^\mu - T\,G_0 \, \mathbf{\Lambda}^\nu  G_0\,\Gamma^\mu\,.
             \end{split}
             \end{equation}
             The essence of this relation is that the object $\Gamma^{\mu\nu}$, just as the vertex $\Gamma^\mu$ itself,
             includes meson bound-state poles in the total quark-antiquark momentum which have their origin in the scattering matrix $T$.
             Consider, for example, the case where the index $\mu$ corresponds to a pion and $\nu$ to a photon.
             In that case, $J^{\mu\nu}$ from Eq.~\eqref{scattering-final} describes pion (virtual) photoproduction,
             and $\Gamma^\mu$ in Eq.~\eqref{Gamma-mu-nu-T} represents the pion bound-state amplitude $\Psi$.
             The lowest-lying meson pole in $T$ corresponds to another pion, and
             at the respective mass shell of the internal pion pole $\Gamma^{\mu\nu}$ becomes
             \begin{equation}
                \Gamma^{\mu\nu} \stackrel{P^2=-M^2}{\longlongrightarrow} -\frac{\Psi\,\conjg{\Psi}}{P^2+M^2}\,G_0 \,\mathbf{\Lambda}^\nu\,G_0\,\Psi \,,
             \end{equation}
             where $-P^2$ denotes here the Mandelstam variable $t$ and $M$ the pion mass.
             Comparison with Eq.~\eqref{current-final} implies that the residue at the pole is just the pion's electromagnetic current:
             \begin{equation}
                \Gamma^{\mu\nu} \stackrel{P^2=-M^2}{\longlongrightarrow}  -\frac{J^\mu \,\Psi}{P^2+M^2}\,.
             \end{equation}
             Thus, the pion's electromagnetic form factor, sketched in Fig.~\ref{fig:preliminary}d,
             will be recovered in the solution of the inhomogeneous BSE~\eqref{ibse-Gamma-mu-nu} and
             enters the pion photoproduction amplitude $J^{\mu\nu}$ through the term $\Gamma^{\mu\nu}$
             which provides a natural description of an 'offshell pion'.
             In addition, all other meson poles whose bound-state amplitudes produce non-zero form factors with a pion and a photon will appear
             in the vertex $\Gamma^{\mu\nu}$ and subsequently in $J^{\mu\nu}$ as well.

        \subsection{Baryons in rainbow-ladder}\label{sec:baryons-in-rl}

\begin{figure*}[t]
                    \begin{center}

                    \includegraphics[scale=0.048]{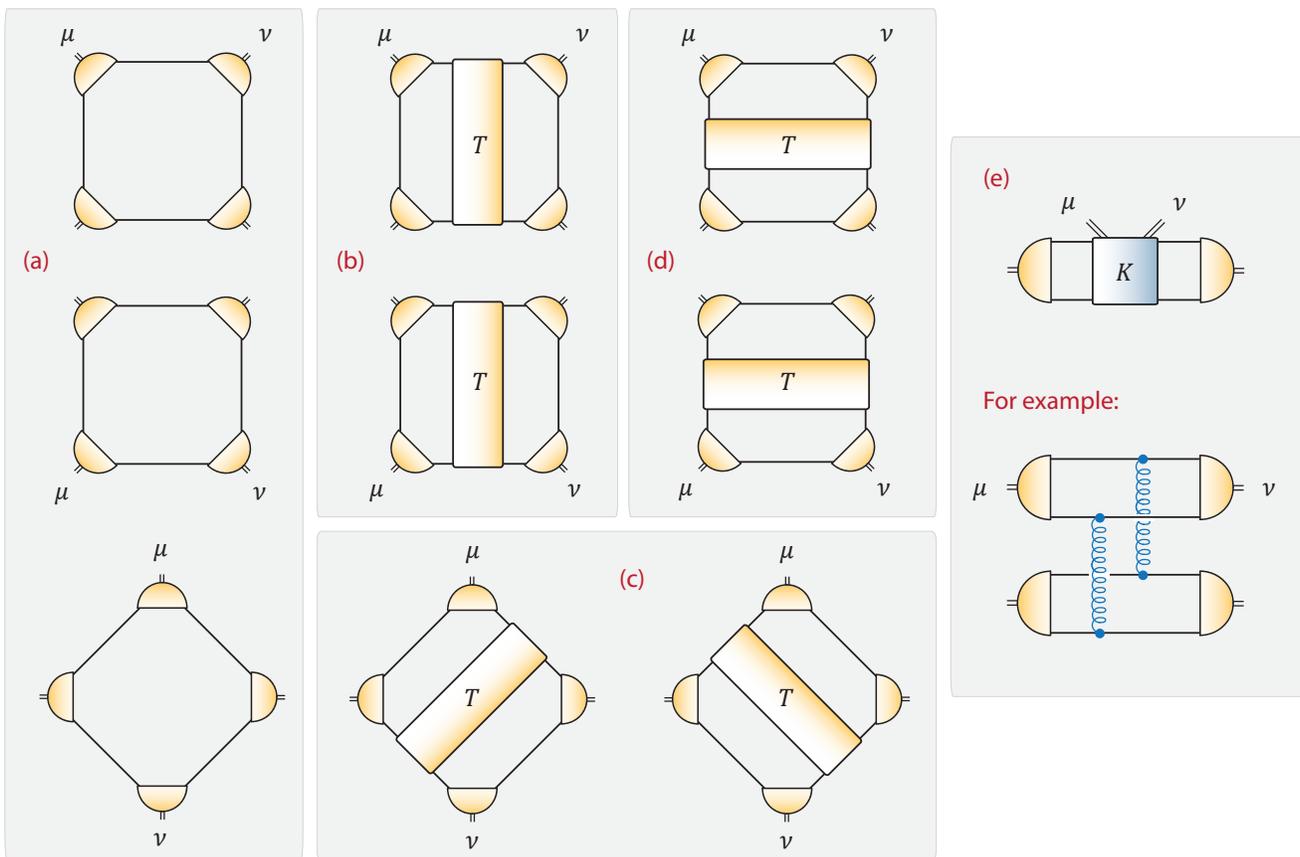}
                    \caption{(Color online) $\pi\pi$ scattering amplitude $J^{\mu\nu}$ in rainbow-ladder truncation, cf.~Eq.~\eqref{J-mu-nu-Meson}.
                                            A further symmetrization in the indices $\mu$ and $\nu$ duplicates the number of diagrams.
                                            To obtain the pion Compton scattering amplitude, replace the amplitudes attached to $\mu$ and $\nu$ by quark-photon vertices.
                                              } \label{fig:pipiscattering}

                    \end{center}
        \end{figure*}

             After having studied the general features of the scattering amplitude,
             we turn to practical applications and analyze its decomposition under specific assumptions for the involved kernels.
             The baryon's bound-state equation~\eqref{boundstate-eq} was recently solved for both nucleon and $\Delta$ baryons
             by neglecting the three-quark irreducible part $K_{(3)}$ that enters the kernel~\eqref{three-body-kernel}
             and implementing a rainbow-ladder truncation for the quark-quark kernel $K_{(2)}$~\cite{Eichmann:2009qa,SanchisAlepuz:2011jn}.
             In our context, the rainbow-ladder truncation implies $K^\mu_{(2)}=0$ which simplifies the structure of $J^{\mu\nu}$ considerably.
             Here, $K_{(2)}$ consists of two bare
             quark-gluon vertices $\sim\gamma^\mu$ that are connected by an
             effective  gluon propagator. In vacuum, a single photon or pion
             cannot couple to a gluon due to Furry's theorem, charge
             conjugation and flavor arguments; however, two photons or pions can.
             To account for that possibility, we retain the term $K^{\mu\nu}_{(2)}$
             in our following considerations. It would describe non-planar processes
             of the type (e) in Fig.~\ref{fig:pipiscattering} in the case of $\pi\pi$ scattering
             and yield analogous diagrams for the scattering on baryons.

             We simplify the notation by writing
             \begin{equation}
                 \Gamma^\mu_a = \underbrace{\Gamma^\mu}_{a} \otimes \underbrace{S^{-1}\otimes S^{-1}}_{bc} \,, \quad
                 \Gamma^\mu_a\,K_a = \underbrace{\Gamma^\mu}_{a} \otimes \underbrace{K_{(2)}}_{bc}\,,
             \end{equation}
             and similarly for $\Gamma^{\mu\nu}_a$ and $K^{\mu\nu}_a$.
             The index $a$ labels the quark which couples to the current,
             or the spectator quark with respect to the two-body kernel,
             and $\{a,b,c\}$ is an even permutation of $\{1,2,3\}$.
             The relations of Eqs.~\eqref{Gamma-mu-3body} and~(\ref{Gamma-mu-nu-3body}--\ref{K-mu-nu-3body})
             reduce in rainbow-ladder truncation to
             \begin{equation}
             \begin{aligned}
                 \mathbf{\Gamma}^\mu &= \sum_a \Gamma_a^\mu\,, &
                 \mathbf{\Gamma}^{\mu\nu} &= \sum_a \Big( \Gamma_a^{\mu\nu} +  \Gamma_b^{\{\mu}\,\Gamma_c^{\nu\}}\Big)\,,\\
                 K^\mu &= \sum_a \Gamma_a^\mu\,K_a\,, &
                 K^{\mu\nu} &= \sum_a \Big( \Gamma_a^{\mu\nu}\,K_a + K_a^{\mu\nu}\Big),
             \end{aligned}
             \end{equation}
             which yields for the quantities $\mathbf{\Lambda}^{\mu}$ and $\mathbf{\Lambda}^{\mu\nu}$ defined in Eq.~\eqref{lambda}:
             \begin{equation}
             \begin{split}
                 \mathbf{\Lambda}^{\mu} &= \sum_a \Gamma_a^\mu\, (\mathds{1}_a-K_a)\,, \\
                 \mathbf{\Lambda}^{\mu\nu} &= \sum_a \Big(\Gamma_a^{\mu\nu}\, (\mathds{1}_a-K_a) +  \Gamma_b^{\{\mu}\,\Gamma_c^{\nu\}} - K_a^{\mu\nu} \Big).
             \end{split}
             \end{equation}
             We can strip the notation to its bare minimum by suppressing all occurrences of
             $G_0 = S\otimes S\otimes S$ as well: all amplitudes, vertices, kernels and T-matrices
             are amputated and their connection via dressed quark propagators is implicit.
             Furthermore, we can absorb the two-quark quantities $(\mathds{1}_a-K_a)$ in the baryon amplitudes and define:
             \begin{equation}\label{hatched}
                 (\mathds{1}_a-K_a)\,\Psi_i =: \Psi_i^a, \quad
                 \conjg{\Psi}_f\,(\mathds{1}_a-K_a) = \conjg{\Psi}_f^a\,.
             \end{equation}
             The baryon current of Eq.~\eqref{current-final} in rainbow-ladder truncation becomes in that condensed notation:
             \begin{equation}
                 J^\mu = \sum_a \conjg{\Psi}_f\,\Gamma^\mu_a\,\Psi_i^a = \sum_a \conjg{\Psi}_f^a\,\Gamma^\mu_a\,\Psi_i\,.
             \end{equation}

             Similarly, to obtain the scattering matrix $J^{\mu\nu}$, we separate $G$ into its disconnected part and the term that involves
             the T-matrix via Eq.~\eqref{T-definition}: $G=1+T$.
             Hadron resonances appear in $T$ only, hence such a separation does not alter the conclusions of Section~\ref{sec:resonances}.
             The scattering amplitude thus becomes
             \begin{equation}
             \begin{split}
                 J^{\mu\nu} &= \sum_{aa'} \conjg{\Psi}_f^a \left[ \Gamma_a^{\{\mu}\,\Gamma_{a'}^{\nu\}} + \Gamma_a^{\{\mu}\,T\,\Gamma_{a'}^{\nu\}} \right] \Psi_i^{a'} \\
                            &- \sum_a \conjg{\Psi}_f \left[ \Gamma_a^{\mu\nu}\,\Psi_i^a + \Gamma_b^{\{\mu}\,\Gamma_{c}^{\nu\}}\,\Psi_i - K_a^{\mu\nu}\,\Psi_i \right],
             \end{split}
             \end{equation}
             which we can rearrange according to the breakdown shown in Fig.~\ref{fig:4-point-function}:
             \begin{equation}\label{J-mu-nu-RL}
             \begin{split}
                 J^{\mu\nu} &= \sum_{a} \conjg{\Psi}_f^a \, \Gamma_a^{\{\mu}\,\Gamma_{a}^{\nu\}} \,\Psi_i^{a} \\
                            &+ \sum_{a\neq a'} \left[ \conjg{\Psi}_f^a \, \Gamma_a^{\{\mu}\,\Gamma_{a'}^{\nu\}} \,\Psi_i^{a'}
                             - \textstyle\frac{1}{2}\,\conjg{\Psi}_f \, \Gamma_a^{\{\mu}\,\Gamma_{a'}^{\nu\}} \,\Psi_i \right]  \\
                            &- \sum_{a} \conjg{\Psi}_f \, \Gamma_a^{\mu\nu} \,\Psi_i^{a} \\
                            &+ \sum_{aa'} \conjg{\Psi}_f^a \, \Gamma_a^{\{\mu}\,T\,\Gamma_{a'}^{\nu\}} \,\Psi_i^{a'} \\
                            &+ \sum_a \conjg{\Psi}_f\,K_a^{\mu\nu}\,\Psi_i\,.
             \end{split}
             \end{equation}
             Here, each line from top to bottom matches one of the diagrams (a)-(e) in Fig.~\ref{fig:4-point-function}.
             The colored amplitudes in the figure correspond to $\Psi_i$ and $\Psi_f$
             whereas the hatched amplitudes represent the quantities $\Psi_i^a$, $\Psi_f^a$ from Eq.~\eqref{hatched}.
             The third line in the above equation is identical to diagram (c)
             since $\Gamma^{\mu\nu}$ from Eq.~\eqref{Gamma-mu-nu-T} simplifies in rainbow-ladder truncation to
             \begin{equation}\label{Gamma-mu-nu-RL}
                 \Gamma^{\mu\nu} = -T\,\big( \Gamma^{\{\mu} \,\Gamma^{\nu\}}\big)\,.
             \end{equation}
             Meson poles in the $t-$channel appear in diagram (c) whereas $s-$ and $u-$channel hadron resonances are contained in diagram (d).
             Handbag contributions appear in all diagrams except type (b);
             one can however show that diagrams (a), (c) and the pure handbag content of (d) can be merged into a single graph of type~(c)
             where the quark-antiquark scattering matrix $T$ is replaced by the Green function $G$.

        \subsection{Mesons in rainbow-ladder}\label{sec:mesons-in-rl}

             The representation of the scattering amplitude $J^{\mu\nu}$ in Eq.~\eqref{scattering-final} is general and holds for baryons and mesons alike.
             In the meson case,
             a rainbow-ladder truncation again removes the gauged quark-antiquark kernel $K^\mu$ from Eqs.~(\ref{current-final}--\ref{scattering-final}).
             Here the notation of the previous section is somewhat impractical and so we use the following abbreviations:
             \begin{equation}
                 \Gamma^\mu_\uparrow = \Gamma^\mu \otimes S^{-1}\,, \quad
                 \Gamma^\mu_\downarrow = S^{-1} \otimes \Gamma^\mu\,,
             \end{equation}
             such that $\mathbf{\Gamma}^\mu = \Gamma^\mu_\uparrow + \Gamma^\mu_\downarrow$, and $\Gamma^{\mu\nu}_{\uparrow\downarrow}$ is defined analogously.
             Suppressing the notation of the propagators, the rainbow-ladder current becomes %$J^\mu$ becomes
             \begin{equation}
                 J^\mu = \conjg{\Psi}_f\,\Big( \Gamma^\mu_\uparrow + \Gamma^\mu_\downarrow \Big)\,\Psi_i\,,
             \end{equation}
             which is just the impulse approximation where the current couples to the upper and lower quark lines.

             In the case of the scattering amplitude $J^{\mu\nu}$, the first term in \eqref{scattering-final} stemming from the disconnected Green function $G_0$ yields
             \begin{equation}
             \begin{split}
                 & \conjg{\Psi}_f\,(\Gamma_\uparrow+\Gamma_\downarrow)^{\{\mu} (\Gamma_\uparrow+\Gamma_\downarrow)^{\nu\}} \Psi_i \\
                 & = \conjg{\Psi}_f\,\Big( \Gamma_\uparrow^{\{\mu}\,\Gamma_\uparrow^{\nu\}} + \Gamma_\downarrow^{\{\mu}\,\Gamma_\downarrow^{\nu\}} + 2\,\Gamma_\uparrow^{\{\mu}\,\Gamma_\downarrow^{\nu\}} \Big)\,\Psi_i
             \end{split}
             \end{equation}
             whereas the contribution from $\mathbf{\Gamma}^{\mu\nu}$ reads
             \begin{equation}
                 -\conjg{\Psi}_f\,\Big( \Gamma_\uparrow^{\mu\nu} + \Gamma_\downarrow^{\mu\nu} + \Gamma_\uparrow^{\{\mu}\,\Gamma_\downarrow^{\nu\}} \Big)\,\Psi_i\,.
             \end{equation}
             The final result,
             \begin{equation}\label{J-mu-nu-Meson}
             \begin{split}
                 J^{\mu\nu} = \conjg{\Psi}_f \,\Big[ & \,\Gamma^{\{\mu}_\uparrow \,\Gamma^{\nu\}}_\uparrow  +
                                                         \Gamma^{\{\mu}_\downarrow \,\Gamma^{\nu\}}_\downarrow +
                                                         \Gamma^{\{\mu}_\uparrow \,\Gamma^{\nu\}}_\downarrow  \\
                                                    & + \Gamma^{\{\mu}_\uparrow\,T\, \Gamma^{\nu\}}_\uparrow + \Gamma^{\{\mu}_\downarrow   \,T\, \Gamma^{\nu\}}_\downarrow \\
                                                    & + \Gamma^{\{\mu}_\uparrow\,T\,\Gamma^{\nu\}}_\downarrow  +\Gamma^{\{\mu}_\downarrow\,T\, \Gamma^{\nu\}}_\uparrow \\
                                                    & - \Gamma^{\mu\nu}_\uparrow - \Gamma^{\mu\nu}_\downarrow   + K^{\mu\nu} \Big]\,\Psi_i\,,
             \end{split}
             \end{equation}
             is illustrated in Fig.~\ref{fig:pipiscattering} in the case of $\pi\pi$ scattering.
             The first line in Eq.~\eqref{J-mu-nu-Meson} corresponds to the impulse-approximation diagrams of type (a);
             the second and third lines represent the diagrams (b) and (c) with a vertical and diagonal T-matrix insertion, respectively;
             and the terms with $\Gamma^{\mu\nu}_{\uparrow\downarrow}$ correspond to the horizontal $t-$channel T-matrix insertion, type (d), stemming from Eq.~\eqref{Gamma-mu-nu-RL}.
             Finally, we have retained the term $K_2^{\mu\nu}$ as well which can appear if the gluon-exchange kernel is resolved into quark loops
             and lead to non-planar terms of type~(e). However, since the rainbow-ladder gluon is a
             model ansatz without explicit quark-loop content such a term does not contribute.

             Fig.~\ref{fig:pipiscattering} demonstrates that the impulse-approximation diagrams alone
             are inconsistent with the rainbow-ladder truncation. In Refs.~\cite{Bicudo:2001jq,Cotanch:2002vj},
             the $\pi\pi$ scattering matrix was computed upon implementing the full set of diagrams (a--d).
             The resulting isospin amplitudes reproduce Weinberg's theorem for the scattering lengths at threshold and the Adler zero in the chiral limit.
             Moreover, they show excellent agreement with chiral perturbation theory in general kinematics, and the $\sigma$ and $\rho-$meson poles in the respective
             $s$, $t$ and $u$ channels are readily recovered.

             In contrast to the baryon case, all diagrams in Fig.~\ref{fig:pipiscattering} except type (e) 
             can be computed fully self-consistently, with comparatively modest numerical effort, if the external particles are identical.
             This can be seen by replacing all occurrences
             of $T$ in Fig.~\ref{fig:pipiscattering} by the Green function $G$ via Eq.~\eqref{T-definition},
             whereby the impulse-approximation diagrams pick up a minus sign.
             $G$ satisfies Dyson's equation~\eqref{Dyson-for-G-1} which, upon contracting with two pion bound-state amplitudes $\Psi$
             in the spirit of Eqs.~(\ref{resonances-bse-1}--\ref{resonances-bse-4}), can be transformed into an inhomogeneous BSE:
             \begin{equation}\label{mesons-bse-for-psi-mu-nu}
             \begin{split}
                 \Psi^{\mu\nu} &:= G_0^{-1}\,G\,(\Psi\,S\,\Psi) \quad \Rightarrow \\
                 \Psi^{\mu\nu} &= \Psi\,S\,\Psi + K\,G_0\,\Psi^{\mu\nu}\,.
             \end{split}
             \end{equation}
             $\Psi^{\mu\nu}$ is a four-point function with two spinor and two pseudoscalar legs
             and thus consists of 8 Dirac basis elements. Since the two pions are onshell,
             it depends on four Lorentz-invariant kinematic variables.
             In fact, the result of Ref.~\cite{Cotanch:2002vj} was obtained by
             a self-consistent solution of Eq.~\eqref{mesons-bse-for-psi-mu-nu}
             and its subsequent implementation in the scattering amplitude $J^{\mu\nu}$.
             We finally note that the same set of diagrams also appears in the photon four-point function that enters
             the hadronic light-by-light contribution to the muon anomalous magnetic moment~\cite{Goecke:2010if}.

        \section{Discussion}\label{sec:discussion}

             The result for the scattering amplitude in Eq.~\eqref{scattering-final} provides,
             in principle, a complete description of scattering processes in QCD
             if a hadron couples to two particles with $q\bar{q}$ quantum numbers.
             Thus, the approach can be applied to a variety of reactions such as nucleon Compton scattering,
             nucleon-pion scattering, meson photo- and electroproduction, $\pi\pi$ scattering,
             or strangeness and charm production processes; but also to
             crossed-channel reactions, for example exclusive
             $p\bar{p}$ annihilation into two photons that will be studied
             with the PANDA experiment at FAIR/GSI~\cite{Lutz:2009ff}.
             The diagrammatic decomposition of the scattering amplitude
             allows to correlate effects at the hadron level in various kinematical limits with the underlying ingredients in QCD.
             The appearance of the $q\bar{q}$ scattering matrix in the $t$ channel and,
             in the case of baryons, the $qqq$ scattering matrix in the $s$ and $u$
             channels induces the existence of bound-state poles at certain values of the invariant Mandelstam variables.
             This feature can be exploited to compute the 'offshell behavior' of the pion electromagnetic form factor
             that appears in pion electroproduction~\cite{Horn:2007ug}.
             Studying virtual Compton scattering, on the other hand,
             enables to investigate two-photon corrections to nucleon form factors~\cite{Arrington:2011dn},
             and isolating the handbag structure in DVCS can be used to draw conclusions about generalized parton distributions.

             The approach relegates the determination of the scattering amplitudes
             to the knowledge of QCD's Green functions which are encoded in
             the quark propagator and the two- and three-quark irreducible kernels.
             The aspiration to describe experimental cross sections
             thus becomes a matter of constructing a truncation of
             the DSEs that captures the physically relevant features.
             Naturally, in a first step one would implement the widely-used
             rainbow-ladder truncation for the computation of scattering amplitudes.
             We have outlined the resulting expressions for scattering on baryons
             and mesons in Sections~\ref{sec:baryons-in-rl} and~\ref{sec:mesons-in-rl}, respectively,
             and they are illustrated in Figs.~\ref{fig:4-point-function}--\ref{fig:pipiscattering}.
             We have already mentioned that a gluon-exchange interaction yields a good description of
             pseudoscalar and vector-meson properties in the light-quark range, e.g., in view of
             $\pi\pi$ scattering~\cite{Cotanch:2002vj}.
             It is, however, also well-suited for a broader range of hadron properties, for example:
             the charmonium and bottomonium region~\cite{Blank:2011ha};
             nucleon, $\Delta$ and $\Omega$ masses~\cite{Eichmann:2009qa,SanchisAlepuz:2011jn};
             the nucleon's electromagnetic, pseudoscalar and axial form factors~\cite{Eichmann:2011vu,EF-Axial};
             or the $\Delta$ electromagnetic and $N\Delta\gamma$ and $N\Delta\pi$ transition
             form factors~\cite{Nicmorus:2010sd,Nicmorus:2010mc,EN-Delta,Mader:2011zf}.
             Judging from those results, its implementation in the framework of scattering processes seems to be well justified.

             Nevertheless, there are important effects which are not captured by a rainbow-ladder interaction.
             One example is the opening of hadronic decay channels such as $\rho\rightarrow\pi\pi$ and $\Delta\rightarrow N\pi$
             that produce non-analyticities in the light pion-mass regime and are essential
             for reproducing the characteristic cut structures in scattering amplitudes.
             A rainbow-ladder kernel does not provide such decay mechanisms but produces instead stable bound states.
             Interactions beyond rainbow-ladder are also vital for the features of scalar,
             axial-vector and isosinglet mesons~\cite{Alkofer:2008et,Chang:2011vu,Fischer:2009jm},
             meson radial excitations~\cite{Krassnigg:2009gd,Blank:2011qk}
             or heavy-light mesons~\cite{Nguyen:2010yh}. This suggests analogous conclusions
             when considering baryon excitations and open strangeness or charm production processes.
             In that respect it will be crucial to complement the computation of scattering amplitudes by a
             simultaneous development of kernels beyond rainbow-ladder truncation.

             Probably the most prominent example for missing contributions in rainbow-ladder
             are pion-cloud corrections in the chiral and low-momentum region.
             Their absence is visible in various hadron properties such as
             nucleon electromagnetic form factors~\cite{Eichmann:2011vu}.
             Dressing the nucleon's 'quark core' by final-state interactions
             with pseudoscalar mesons is among the goals pursued in the approaches listed in the introduction,
             such as for example by the EBAC group at JLAB~\cite{Matsuyama:2006rp}.
             In the Dyson-Schwinger framework, pion-cloud effects should ultimately be implemented at the level of QCD's Green functions.
             From a microscopic point of view, pion exchange corresponds to a gluon resummation.
             This allows to reconstruct the relevant topologies for generating phenomenological
             chiral cloud effects from the underlying dynamics in QCD.
             Pion-loop contributions will thus appear in all Green functions
             with non-vanishing quark content and lead to complicated gluon-interaction diagrams.
             A more manageable route was taken in Refs.~\cite{Fischer:2007ze,Fischer:2008wy,Goecke:2010if},
             where the assumption that the quark-antiquark T-matrix in the low-energy region
             is dominated by pion exchange was used to derive a consistent % leads to a consistent
             truncation of the DSEs beyond rainbow-ladder.
             Employing the resulting $q\bar{q}$ kernel in meson Bethe-Salpeter studies
             recovers desired pion-cloud induced features, e.g., a reduction of meson masses in the low quark-mass region~\cite{Fischer:2008wy}.
             The eventual implementation of such hybrid kernels with quark, gluon and effective pion degrees of freedom in the baryon sector
             would mark a notable step forward.

             The truncation dependence of the Dyson-Schwinger approach finally provides also opportunities.
             The construction of Eq.~\eqref{scattering-final} allows to trace experimentally observable effects
             at the hadron level back to the underlying Green functions of QCD
             and can thus supply information and potential constraints on their properties.
             In that respect we emphasize that it is not only the behavior of the quark mass function
             that has ramifications at the hadron level. The quark propagator in Landau gauge is well determined
             from lattice QCD~\cite{Bowman:2005vx,Kamleh:2007ud},
             and even the rainbow-ladder truncated Dyson-Schwinger result for the propagator agrees reasonably well
             with those data. Equally important for hadron properties
             is the structure of the two-quark and three-quark kernels and the quark-gluon vertex upon which they depend.
             Examples are the question of the dominance of two- or three-quark interactions in the baryon's excitation spectrum, %~\cite{a},
             or the insufficiencies of a gluon-exchange interaction for the observables mentioned before.
             Putting constraints from experiment on these (however gauge-dependent) quantities would be indeed desirable.

             Apart from the truncation dependence there are
             further practical difficulties in computing Eq.~\eqref{scattering-final}.
             One of them was already mentioned in Section~\ref{sec:resonances}
             for the baryon case and concerns the computational effort in accessing the $s-$channel
             structure in the scattering amplitude, which is the nucleon resonance region, as it involves the full three-quark scattering matrix.
             Another problem is limited kinematical access to the phase space of such processes. This is due to the singularity
             structure of the quark propagator that inevitably emerges in the solution of the quark DSE as a consequence of analyticity.
             While not a problem in principle, present numerical implementations only sample
             the kinematically safe momentum ranges of the quark propagators that appear inside a momentum loop.
             This leads to limitations, e.g., when computing masses of excited hadrons,
             and restricts the kinematical access to hadron form factors and scattering amplitudes.
             For instance, accessing the nucleon resonance region above threshold in the $N\gamma$ or $N\pi$ scattering amplitudes
             might require an extrapolation from spacelike photon or pion momenta.

             These considerations make clear that the computation of scattering amplitudes can only happen in gradual steps.
             The primary emphasis would be put on understanding the mechanisms in QCD that determine
             the behavior of the partial-wave amplitudes in certain kinematical regions,
             or to analyze and disentangle resonant and non-resonant
             background effects in the $s$, $t$ and $u$ channels and study offshell effects.
             One can furthermore isolate the properties of the quark core by
             investigating the current-mass dependence of the results,
             in particular in the region where hadrons become stable bound states and the pion cloud is suppressed.
             Eventually, with advanced numerics, more sophisticated interactions and improved kinematical coverage,
             a systematic description of scattering processes from the underlying dynamics in QCD seems certainly possible.

        \section{Conclusions and outlook}\label{sec:conclusion}

             We have detailed a systematic construction to describe scattering processes
             of photons and mesons with hadrons in the Dyson-Schwinger framework of QCD.
             The approach is Poincar\'e-covariant and covers both perturbative and
             nonperturbative momentum regions as well as the full quark mass range.
             Hadronic scattering amplitudes are resolved in terms of the underlying dynamics
             in QCD by coupling the external currents to all dressed quarks.
             Electromagnetic gauge invariance and crossing symmetry are therefore satisfied by construction.
             The amplitudes can be computed consistently if their microscopic ingredients, i.e., the dressed quark propagator and the
             irreducible two- and three-quark kernels, have been determined in advance.
             The resulting expressions allow to relate hadron resonances as well as
             background effects which appear in the $s$, $t$ and $u$ channels to the structure
             of the internal two- and three-quark scattering matrices.

             Potential applications cover a wide range of reactions such as
             Compton scattering, nucleon-pion scattering, meson photo- and electroproduction
             or exclusive proton-antiproton annihilation processes. Such studies would
             further allow to access offshell effects and two-photon contributions in the extraction
             of form factors, and they can potentially also provide information on generalized parton distributions.

             We have discussed the practical feasibility of the approach and found that,
             while scattering on mesons is comparatively unproblematic from a computational point of view,
             the application in the baryon sector will require substantial numerical effort.
             This is especially true for the nucleon resonance region,
             whereas $t-$channel effects are easier to access.
             We have exemplified the approach in a rainbow-ladder truncation,
             where the two-quark kernel is represented by a dressed gluon exchange and the three-quark kernel is omitted.
             Its successful application in various studies of meson and baryon phenomenology
             suggests that this setup provides a suitable starting point to study hadron-meson and hadron-photon reactions as well.
             Ultimately, in view of implementing pion-cloud corrections and hadronic decay channels,
             the computation of scattering processes should be paralleled by an effort to go beyond rainbow-ladder.
             The derivation that we have presented is general and can accommodate such purposes.

        \section*{Acknowledgments}

            We would like to thank R.~Alkofer, M.~Blank, T.~G\"ocke, A.~Krassnigg and R.~Williams
            for valuable discussions. This work was supported by the Austrian Science Fund FWF under
            Erwin-Schr\"odinger-Stipendium No.~J3039,
            the Helmholtz International Center for FAIR
            within the LOEWE program of the State of Hesse
            and the Helmholtz Young Investigator Group No.~VH-NG-332.

%%%%%%%%%%%%%%%%%%%%%%%%%%%%%%%%%%%%%%%%%%%%%%%%%%%%%%%%%%%%%%%%%%%%%%%%%%%%%%%%%%%%%%%%%%%%%%%%%%%%%%%%%%%%%%%%%%%%%%%%%%%%%%%%%%%%%%%%%%%%%%%%%%%%%%%%%%%%%%%%%%%%%%%%%%%%%%%
%%%%%%%%%%%%%%%%%%%%%%%%%%%%%%%%%%%%%%%%%%%%%%%%%%%%%%%%%%%%%%%%%%%%%%%%%%%%%%%%%%%%%%%%%%%%%%%%%%%%%%%%%%%%%%%%%%%%%%%%%%%%%%%%%%%%%%%%%%%%%%%%%%%%%%%%%%%%%%%%%%%%%%%%%%%%%%%

\bibliographystyle{apsrev4-1-mod}

\bibliography{lit-scattering}

\end{document}